\documentclass[showpacs]{revtex4}
\usepackage{epsfig}
\usepackage{times}
\usepackage{enumerate}
\usepackage{amsmath}
\usepackage{graphicx}
\usepackage{wasysym}

\begin{document}
\title{Collective foraging in heterogeneous landscapes}

\author{Kunal Bhattacharya$^{1,}$\footnote{Corresponding author:\\Email: kunalb@pilani.bits-pilani.ac.in} and Tam\'as Vicsek$^{2}$}
\affiliation
{$^1$Department of Physics, Birla Institute of Technology and Science, Pilani 333031, India\\
$^2$Department of Biological Physics, E\"otv\"os University Budapest, H-1117, Hungary}

\begin{abstract}
Animals foraging alone are hypothesized to optimize the encounter rates with resources through L\'evy walks. However, the issue of how the interactions between multiple foragers influence their search efficiency is still not completely understood. To address this, we consider a model to study the optimal strategy for a group of foragers searching for targets distributed heterogeneously. In our model foragers move on a square lattice containing immobile but regenerative targets. At any instant a forager is able to detect only those targets that happen to be in the same site. However, we allow the foragers to have information about the state of other foragers. A forager who has not detected any target walks towards the nearest location, where another forager has detected a target, with a probability $\exp{\left(-\alpha d\right)}$, where $d$ is the distance and  $\alpha$ is a parameter. The model reveals that neither overcrowding ($\alpha\to 0$) nor independent searching ($\alpha\to\infty$) is beneficial for the group. For patchy distribution of targets the efficiency is maximum for intermediate values of $\alpha$. Also, in the limit $\alpha\to 0$, the length of the walks can become scale-free.      
\end{abstract}

\maketitle

\section{Introduction}

Many organisms collectively gather resources for survival. For a lone animal looking for sources of food, the foraging efficiency is decided by its searching strategy and the distribution of the targets \cite{levy-opt,oft}. However, when others are around the interactions between them may become important in determining the behaviour of the group as a whole. For example, an animal which is unsuccessful in locating food by searching on its own, can instead locate other distant members who are successful and join them. Such behaviour is well documented across different species \cite{fish,sea-bird,optimal}. However, what would be the ideal nature of such interaction being most advantageous for the group? Traditionally, this problem has been studied in behavioural ecology under the paradigms of information sharing and producer-scrounger games \cite{optimal,vickery}. The information sharing theory, in its canonical form, considers foragers simultaneously searching for food as well as looking for opportunities to join others. In the producer-scrounger games the simultaneous execution of the above activities is not possible.

With the advancement of different data logging techniques in animal experiments \cite{levy-data,sims} it has been possible to record the trajectories of animals searching for food. The information from such experiments have allowed for detailed statistical analysis and have helped in shedding light on the possible relationship between movement patterns and efficiency of search. The emergence of heavy tails in the walk-length distributions have prompted physicists to explain the foraging movement of animals in terms of L\'evy walks and flights \cite{foraging-book-1}. Such motion of foragers for specific distribution of the resources is supposed to optimize the rates of the encounters with food items \cite{levy-opt}. 

In general, the success with which a group forages and the observed patterns in the movement of foragers depends on several factors like, the ratio of the amount of resources available to the number of foragers, the nature of forager-forager interaction, the nature of distribution of resources, forager-target interaction and the strategy adopted by foragers. Several field studies and theoretical models have investigated the importance of one or more of these factors. The producer-scrounger paradigm was used to model the collective behaviour of foragers \cite{beauchamp-main,tania}. Cooperation in foraging was shown to develop in a stochastic environment \cite{couzin}. Explanation for observed scale-free move lengths observed in spider monkeys \cite{spider-monkey-exp} have been based on their social behaviour \cite{monkey-model}. Such mobility patterns of foragers was modelled in \cite{zhou} with the agents having unbounded velocities. The role of information transfer in the recruitment of foragers have been studied in colonies of social insects \cite{honeybee,honeybee-model,ant}. Very recently, the influence of communication on the foraging pattern of gazelles was investigated \cite{gazelle}. The study showed that communication over intermediate length scales result in faster searches.       

In this paper we develop a minimal model of collective foraging where the motion of the individual foragers would be random in the absence of any interaction. We investigate the effect of interaction on the search efficiency and the spatio-temporal scaling that emerges in the movement of foragers. The foragers are walkers on a two dimensional square lattice while targets are immobile but regenerative. The foragers have local information about the distribution of the targets. A forager independently searching for targets executes a random walk (RW) on the lattice. However, the foragers possess global information about the state of other foragers, {\it i.e.}, whether they have been successful in locating targets. We characterize the interaction between foragers through the parameter $\alpha$. This parameter controls the propensity of a forager, who is unsuccessful in finding a target in its locality, to approach the nearest site where targets have been discovered by others. We call such a movement, a targeted walk (TW).

We expect that when targets regenerate in the absence of any spatial correlations, the aggregation of foragers would provide no additional advantage as compared to independent searching. This aspect becomes clear from our results. However, when the reappearance of the targets is guided by spatial correlations which result in spatially heterogeneous distribution of the targets in patches \cite{oft,patch-2} we find that the efficiency attains a maximum for intermediate values of $\alpha$. This implies that for foragers who are momentarily unsuccessful in identifying targets, sometimes joining others who are successful, can be beneficial. This is the main result of our paper. Also, the fact that we obtain scale-free TWs under certain conditions suggests collective behaviour as one of the possible mechanisms responsible for the emergence of L\'evy walks in foraging patterns found in the real-world. 

\section{The Model}
\label{def}

Below we describe our model in all details. We consider a two dimensional square lattice of size $L$ with periodic boundary conditions. There are   $N_F$ foragers which are initially distributed randomly on this lattice. The targets (food) are regenerative such that at every time step the total (amount) is $N_T$. We define the neighbourhood of a forager as a circular region of radius $R$ centred on it. We assume that at any instant of time, a forager is able to determine which are the foragers in its neighbourhood that have been successful in identifying and consuming food. However, a forager is only able to detect targets at the site where it has arrived. Let $(x^t_i,y^t_i)$ be the position of a forager $i$ on the lattice at time $t$ and let $j$ be the nearest of the all the foragers in the neighbourhood which have consumed one target at time $t$. The rules governing the behaviour of $i$ are the following: 
\begin{enumerate}[(i)]
\item If there are targets at $(x^t_i,y^t_i)$ then the forager stays at that site and consumes one. Let $h(x^t,y^t)$ denote the number of targets at a site $(x,y)$ at time $t$. Thus, $h(x^{t+1}_i,y^{t+1}_i)=h(x^t_i,y^t_i)-1$ and $(x^{t+1}_i,y^{t+1}_i)=(x^t_i,y^t_i)$. If at any instant of time the number of foragers at any site exceeds the number of targets available, then the targets are distributed randomly between the foragers.
\item If the site $(x^t_i,y^t_i)$ is empty with the forager having arrived at this site by taking a {\it random step} (defined below), or the forager being there in presence of food (at time $t-1$) then the movement of the forager is decided according to the following rules. 
\begin{enumerate}[(a)]
\item If there is no successful forager in the neighbourhood then the forager takes a random step, that is, $(x^{t+1}_i,y^{t+1}_i)$  becomes equal to one of the nearest neighbour coordinates $(x^t_i\pm 1,y^t_i)$ or $(x^t_i,y^t_i\pm 1)$ with equal probability. 
\item If there is a successful forager $j$ in the neighbourhood then with a probability $p_i$ the forager `moves' in the direction of the former and with a probability $1-p_i$ moves out by taking a random step. Here, $p_i=\exp(-\alpha d_{ij})$, where $d_{ij}$ is the Euclidean distance between  
$(x^t_i,y^t_i)$ and $(x^t_j,y^t_j)$. The movement in the direction of $j$ implies moving to a nearest neighbour site such that the distance between $(x^{t+1}_i,y^{t+1}_i)$ and $(x^t_j,y^t_j)$ becomes less than $d_{ij}$. We term this movement as a {\it targeted step} in contrast to a random step and let the forager remember this distance as ${\tilde{d}}^t_i=d_{ij}$. 
\end{enumerate}
\item If the site $(x^t_i,y^t_i)$ is empty and if the forager has arrived at this site by taking a targeted step then the movement of the forager is decided as follows. 
\begin{enumerate}
\item Same as (ii)(a).
\item If there is a forager $j$ in the neighbourhood and $d_{ij} < {\tilde{d}}^{t-1}_i$ then the forager takes targeted step towards $j$ and ${\tilde{d}}^t_i=d_{ij}$.

\item If there is a forager $j$ in the neighbourhood and $d_{ij}\geq {\tilde{d}}^{t-1}_i$ then same as (ii)(b).
\end{enumerate}

\item This rule pertains to the regeneration of targets at every instant. Let $\Delta N(t)$ be the number of targets that have been consumed at any instant of time $t$. We randomly select one of the remaining $N_T-\Delta N(t)$ targets. At a distance $d$ ($1\le d\le L/\sqrt{2}$) from the location of this target a new target is placed, in a random direction. The distance $d$ is chosen from a distribution $P(d)\sim d^{-\gamma}$, where $\gamma$ is a parameter. This process is repeated another $\Delta N(t)-1$ times. Thus, the total number of targets at the beginning of time step $t+1$ again becomes $N_T$. 
\end{enumerate}

We choose such a scheme for the regeneration of targets with a two-fold view. Firstly, it ensures that at any instant the distribution of resources is heterogeneous. The power-law form for $P(d)$ generates targets distributed in well separated clusters. A similar scheme was used to model the spacings between marine food patches in \citep{sims}. Secondly, it provides the scope of simulating the foraging activity as a steady state process in time. A more realistic model would allow the simulation in an infinite lattice where the patches need not regenerate at every time step. In effect our model provides a method to produce a foraging environment that is highly variable in space and time \cite{couzin,monkey-model-2,hierarchical-patch}. 

We call a consecutive sequence of random steps a RW and a consecutive sequence of targeted steps a TW. A forager which searches independently executes a RW until it encounters a target. However, there is a finite probability that such a forager may decide moving towards sites where foragers have already encountered food assuming that targets are clustered in space. It is natural that this probability is distance dependent as there is an energy cost associated with the movement of a forager \cite{cost1}. Also, travelling  over large distances take more time during which the targeted patch becomes increasingly exploited \cite{cost2}. Therefore, nearer the location of the sites, larger is the probability of relocation. This fact is quantified through the definition of $p$ in (ii)(b). In the limit $\alpha\to\infty$ the trajectories of the foragers are essentially RWs. The tendency of foragers to travel towards distant locations, where targets have been discovered, increases with decrease in the value of $\alpha$. Thus, trajectories become mixture of RWs and TWs.   

Note that a TW, when initiated, is not aimed at a particular individual but towards a particular site. The variable ${\tilde{d}}^t$ ensures  that forager has a memory of the distance to the targeted site and therefore the instantaneous movement of the forager is in a direction so as to minimize this distance. On appearance of another site, where food gets detected, closer than the original, the TW gets directed towards it and ${\tilde{d}}^t$ is accordingly modified. In case the last targeted site is exhausted and some feeding site in the neighbourhood gets detected at a distance larger compared to the distance to the former then the forager probabilistically adopts that target depending on the distance. This rule accounts for the fact that when forager that has adopted a strategy of moving towards targets detected by others, will not easily abandon it and return to searching independently. Also, a forager actually approaching a cluster of sites should not get deterred if one of the sites gets emptied. A model with somewhat similar but rather deterministic rules for the movement of the foragers was considered in \cite{gurney} where the effect of group size on foraging rate was studied. The randomness in our model is naturally built in by the probabilistic choice of the step directions of the foragers and the scheme used for the regrowth of the targets.  

In the present study we assume that $R$ is larger than the system size, {\it i.e.}, at any instant, a forager is informed about state of all other foragers. However, it is obvious that the actual use of this information is manifested through the distance dependent probability $p$. Also, we take $N_F=N_T=N$ which would correspond to the case of targets being rather scarce at the same time this equality also ensures that the maximum possible value of targets encountered per unit time per forager is unity. 

\begin{figure*}
\centerline{\epsfig{file=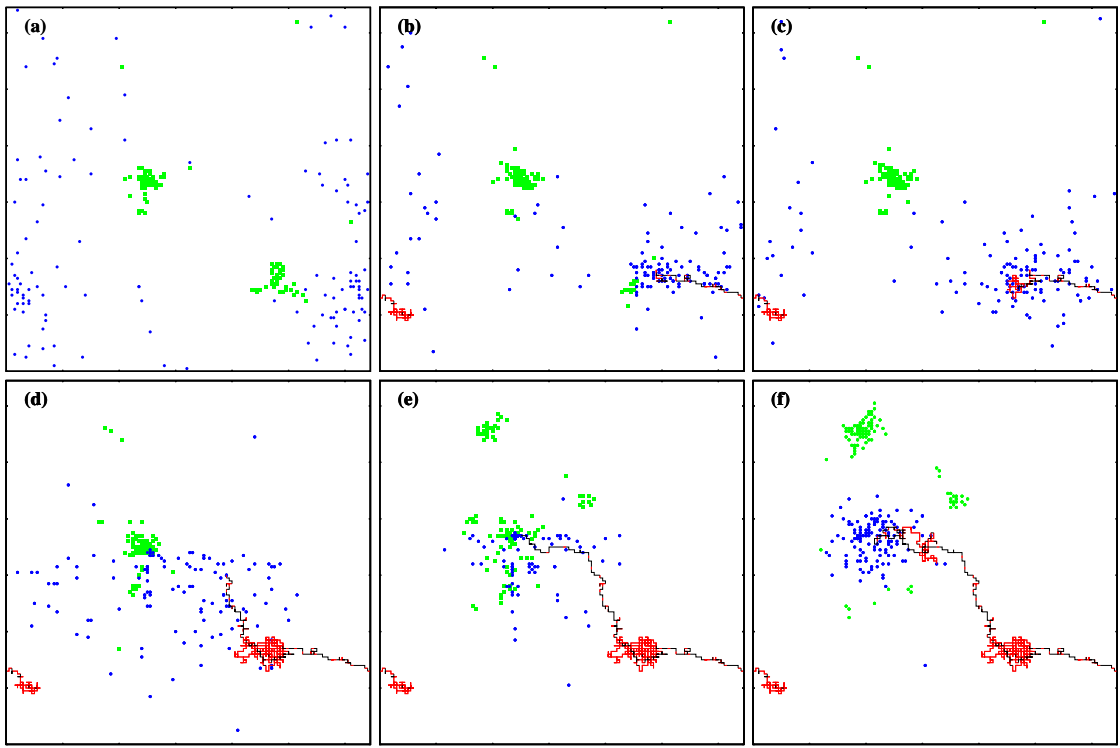,width=15cm}}
\caption{The figure shows the snapshots during evolution of the model for $N=128$, $L=128$, $\gamma=2.5$ and $\alpha=0.0$ at times $t=0$, $194$, $281$, $747$, $802$ and $979$. The foragers are marked with blue circles and the targets are marked with green squares. The path of a typical forager is drawn. The red steps belong to random walks and the black steps belong to targeted walks. The fact that a single lattice site may be multiply occupied by foragers or targets is not separately colour coded.}
\label{fig-1}
\end{figure*}

\section{Results}
\label{results}

In general, not only the regeneration process but the interaction with the foragers is supposed to influence the distribution of targets at long times. However, with the parameters used in this paper, our simulations reveal that the heterogeneity in the spatial distribution of targets at all times is qualitatively independent of $\alpha$ and is controlled through $\gamma$. The larger the value of $\gamma$ the more clustered the targets become and the power-law form for $P(d)$ resulting in rare but very large values of $d$, gives rise to large separation between clusters. 

\begin{figure*}
\centerline{\epsfig{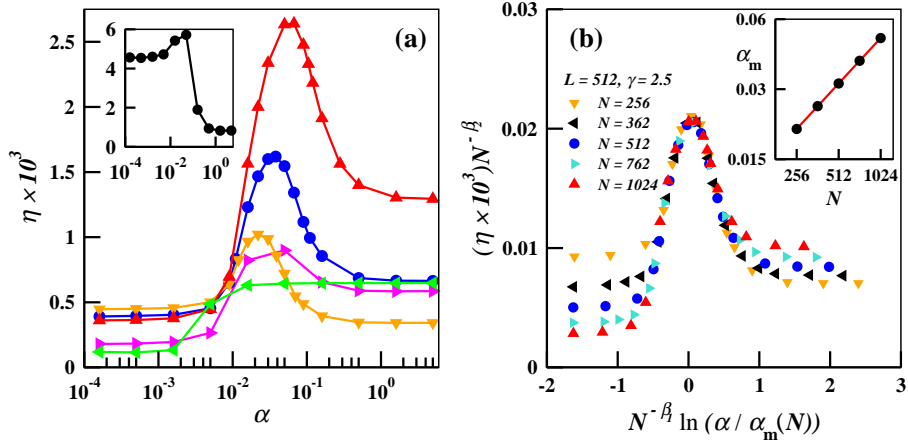}}
\caption{(a) Variation of the efficiency, $\eta$, with $\alpha$ for $L=512$. The different colours used are for $N=1024$, $\gamma=2.5$ (red); $N=512$, $\gamma=2.5$ (blue); $N=256$, $\gamma=2.5$ (orange); $N=512$, $\gamma=2.0$ (magenta) and $N=512$, random regeneration of targets (green). The inset of (a) corresponds to $N=512$ and $\gamma=3.5$. (b) Scaling collapse of $\eta$ for different values of $N$ with $\gamma=2.5$ and $L=512$. The collapse results with $\beta_1=0.15$ and $\beta_2=0.70$. The inset of (b) shows the dependence of $\alpha_m$ on $N$ in a log-log plot. The straight line in red, having slope $\zeta=0.65$, shows the power-law nature of the variation.}
\label{fig-2}
\end{figure*}

In the figure 1 we show the time evolution of the model for the parameters $N=128$, $L=128$, $\gamma=2.5$ and $\alpha=0.0$. The dynamics is deterministic to a certain extent because $\alpha$ is zero. Initially, (Fig.~\ref{fig-1}(a)) the targets appear to be distributed in two well-separated clusters and the foragers are mostly away from these targets. At a later time (Fig.~\ref{fig-1}(b)) the foragers are found to have aggregated over one of the clusters. This happens as soon as one of the foragers is successful in detecting a target belonging to that cluster. The trajectory of a typical foragers is shown. Once the targets in the region, where the foragers are present, gets depleted there is a searching phase (Fig.~\ref{fig-1}(c)). The trajectory of the foragers at this stage are mostly RWs. The search ends as soon as a site in the second cluster is detected by any forager and all the others relocate towards the site. The trajectory during this relocation mostly comprises of TWs (Fig.~\ref{fig-1}(d) and (e)). Eventually, the second cluster is also consumed and another searching phase follows (Fig.~\ref{fig-1}(f)). The simultaneous development of clusters of targets in other regions of the lattice is also visible in the figure.             

We assume that the cost involved in foraging is proportional to the distance travelled by the foragers. The efficiency of searching \cite{levy-opt}, defined as the ratio of the total number of targets consumed to the total distance travelled by all the foragers, is given by
\begin{equation}
\eta=\left\langle\frac{\sum_{t=1}^{\mathcal{T}}\Delta N(t)}{\sum_{t=1}^{\mathcal{T}}N_\textrm{W}(t)}\right\rangle,
\end{equation}
where, $N_\textrm{W}(t)$ is the number of walkers at time $t$. The statistics is collected in the stationary state for $\mathcal{T}=10^6$ time steps in each configuration and $\langle\cdots\rangle$ denotes the average over $200$ configurations. In Fig.~\ref{fig-2}(a) we plot $\eta$ as a function of $\alpha$. For $N=512$ and $L=512$ we are able to compare the efficiency of the searches for targets distributions at different degrees of patchiness characterized by $\gamma$. When targets regenerate randomly across the lattice we find $\eta$ to increase with $\alpha$ and then saturate to a value in the limit of $\alpha\to\infty$. Thus, searching independently is overall beneficial. However, for values of $\gamma=2.0$, $2.5$ and $3.5$, the efficiency is found to be maximum for different values of $\alpha$, respectively. When targets are clustered in space we expect that a forager travelling to a region where targets have already been discovered increases its own chance of encountering a target. However, indiscriminately taking such decisions may not be beneficial. Travelling to distant clusters increases the cost and by the time the forager reaches the region, it is depleted of targets. Therefore, joining others when clusters are discovered nearby and opting to search independently, otherwise is found to be most efficient. Unlike the scenario described in Fig.~\ref{fig-1} this intermediate strategy may amount to the simultaneous discovery and exploitation of more than one cluster or faster discovery of newly generated clusters. For extremely clustered distributions ($\gamma=3.5$) the efficiency in the limit $\alpha\to0$ is larger than that of the $\alpha\to\infty$ limit unlike for lesser values of $\gamma$. This is because the regenerated clusters appear very close to each other and as such aggregation allows the foragers to move from one cluster to another with least amount of exploration. A comparison between different values of $N=256$, $512$ and $1024$ shows that the maximum value of efficiency is larger for larger values of $N$. We quantify this effect by scaling collapse of $\eta$ versus $\alpha$ plots for  five different values of $N$ in Fig.~\ref{fig-2}(b) with $L=512$ and $\gamma=2.5$. The collapse reveals that the width of the maximum in $\eta$ scales as $N^{\beta_1}$, where $\beta_1=0.15$ and the maximum value of the efficiency, $\eta_m$ scales as $N^{\beta_2}$ with $\beta_2=0.70$. We define $\alpha_m$ as the value of $\alpha$ for which $\eta=\eta_m$. The values $\alpha_m$ for which the collapse becomes possible is plotted against $N$ in the inset of Fig.~\ref{fig-2}(b). The plot shows that $\alpha_m\sim N^{\zeta}$ with $\zeta=0.65$. 

\begin{figure*}
\centerline{\epsfig{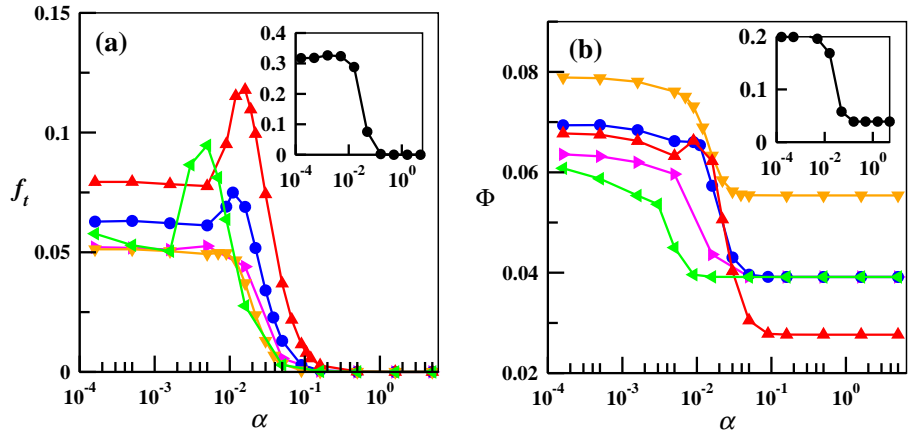}}
\caption{Plots of the fraction of walkers executing targeted walks at any instant, $f_t$ (a) and the flocking order parameter, $\Phi$ (b) against $\alpha$ for $L=512$. The different parameter sets denoted by the different colours in the main areas and the insets are the same as that in the Fig~\ref{fig-2}(a).}
\label{fig-3}
\end{figure*}

\begin{figure*}
\centerline{\epsfig{file=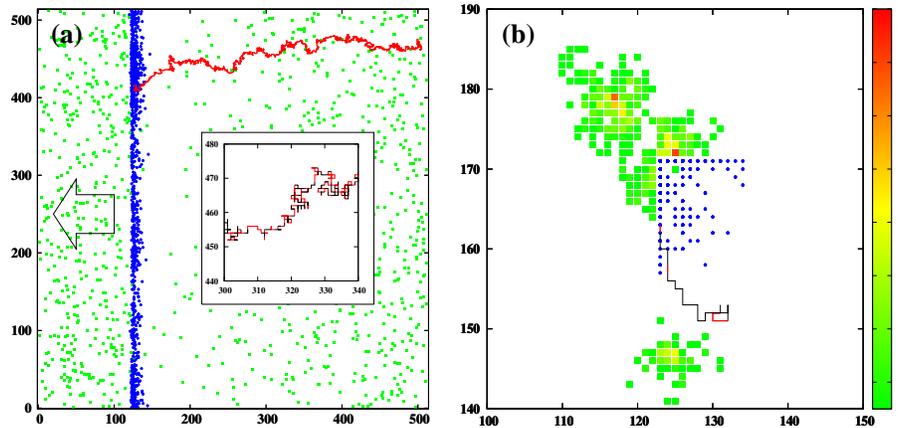,width=12cm}}
\caption{Snapshots of two different patterns exhibited by the model corresponding to regrowth rules for targets and for the parameters $N=512$, $L=512$ and $\alpha=0.0$. A stable moving band observed when targets regenerate randomly across the lattice (a); and wedge formation for $\gamma=3.5$ (b) (only the section of the lattice where foragers and targets are concentrated is shown). In (a) the foragers are marked with blue and the targets are marked with green. The path of typical forager is marked with red. The arrow points to the direction of motion of the band. The multiple occupation of sites by foragers or targets is not colour coded. The inset shows a section of the trajectory where the RWs appear in red and TWs in black. In (b) the foragers are marked with blue (multiple occupation ignored) and the path of a typical forager is shown with a color scheme similar to the inset of (a). A colouring scheme for the multiple occupation of sites by targets is used (provided in legend).}
\label{fig-4}
\end{figure*}

Interestingly, near the maximum of efficiency we find that there is a maximum for the fraction of walkers executing TWs. In Fig.~\ref{fig-3}(a) we plot the
averaged quantity, $f_t$, defined as the ratio of the number of foragers executing TWs to the total number of walkers at any instant. As expected $f_t$ goes to zero as $\alpha\to\infty$ since the foragers do not follow each other.
Also, $f_t$ is large as $\alpha\to0$. The activity of the foragers in this limit is similar to the scenario illustrated in Fig.~\ref{fig-1}. The dominant activity is searching (through RWs) in-between the discovery of clusters and the simultaneous discovery of clusters are rare. The TWs only occur in short bouts. For intermediate values of $\alpha$, when there is a possibility that a cluster can be detected while another one is being exploited, the TWs take place more frequently. This results in $f_t$ having a maximum. When targets regenerate randomly across the lattice the TWs are not beneficial in terms of the efficiency as already seen from Fig~\ref{fig-2}(a). Decrease in the value of $\alpha$ favours the increase in aggregation of the foragers through TWs whereas the aggregation reduces the number of forager-target encounters. This competition results in the maximum of $f_t$ for random regeneration. The effect of random reappearance of targets coupled to extreme aggregation for $\alpha\to 0$ and the fact that we use PBC, gives rise to a stable moving band in the steady state as shown in Fig.~\ref{fig-4}(a). The band travels parallel to either of the axis. 

\begin{figure}
\centerline{\epsfig{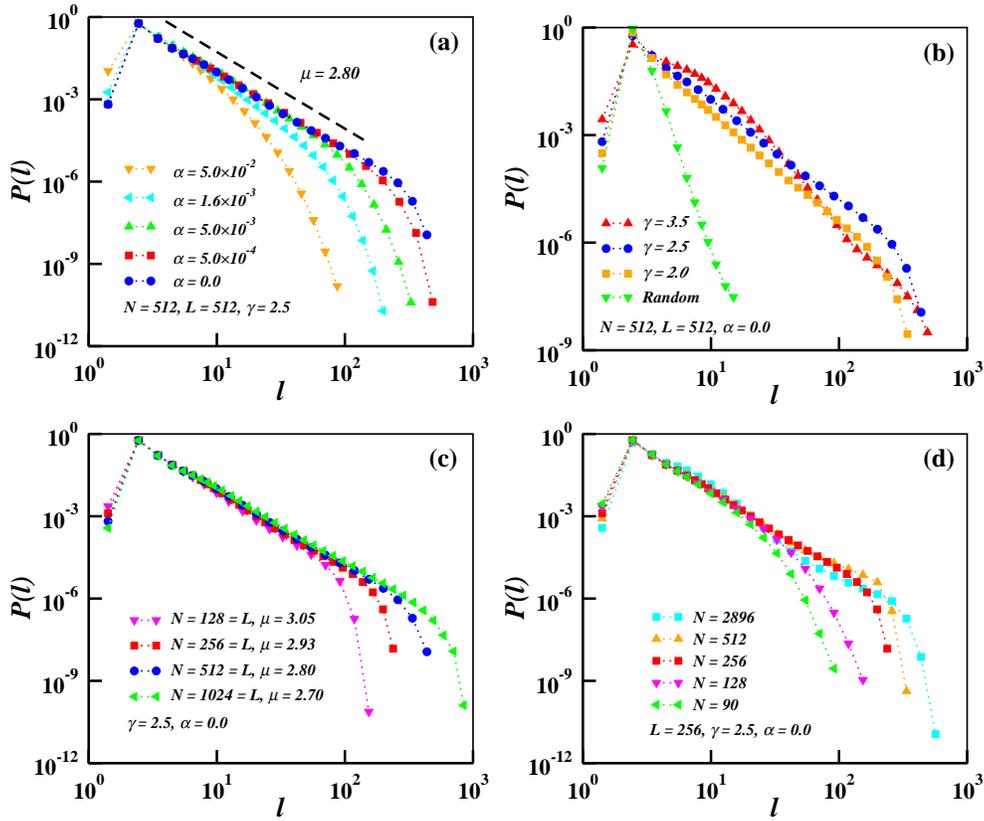}}
\caption{The plots of probability distribution, $P(l)$, of the length of targeted walks. The figures (a), (b), (c) and (d) correspond to the different sets of values of $N$, $L$, $\gamma$ and $\alpha$, as indicated in the legends. The dashed line in (a) is a guide to the eye and indicates the power-law nature in the region for the curve with $N=512$, $L=512$, $\gamma=2.5$ and $\alpha=0.0$. Regression fit with $P(l)\sim l^{-\mu}$ gives the value of $\mu=2.80$. In (c) the values of $\mu$ corresponding to different values of $N(=L)$ resulting from regression fits are provided in legend.}
\label{fig-5}
\end{figure} 
 
\begin{figure}
\centerline{\epsfig{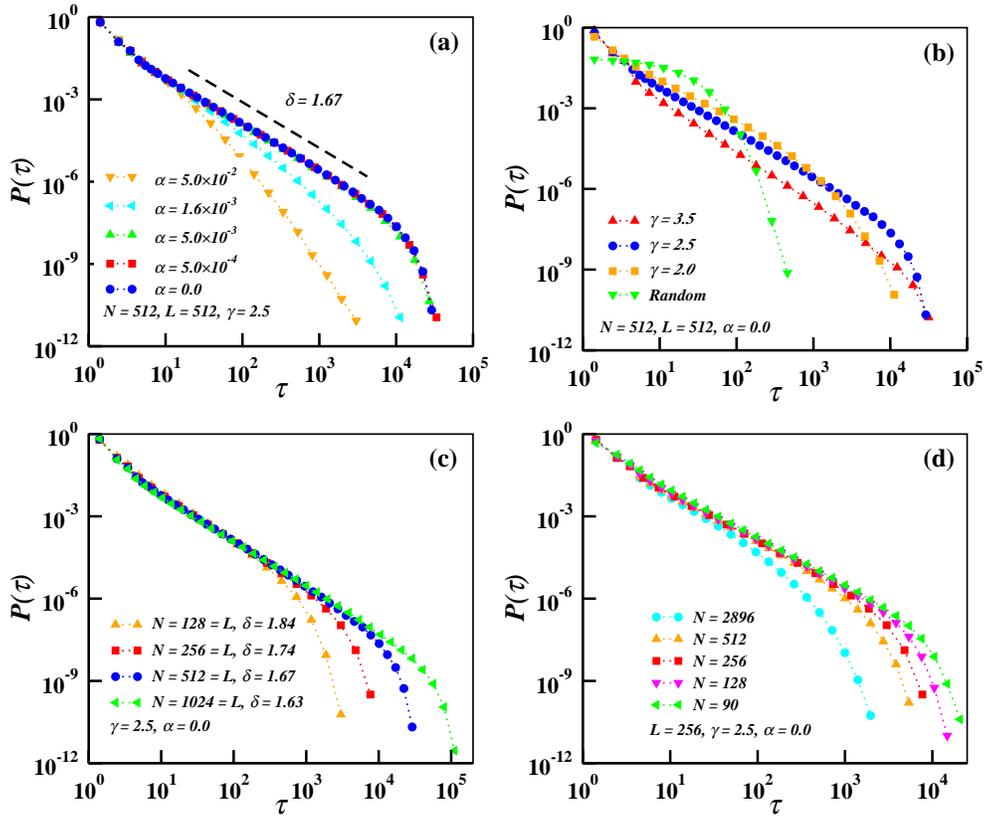}}
\caption{The plots of probability distribution, $P(\tau)$, of waiting times between forager-target encounters. The figures (a), (b), (c) and (d) correspond to different sets of values of $N$, $L$, $\gamma$ and $\alpha$, as indicated in the legends. The dashed line in (a) is a guide to the eye and indicates the power-law nature in the region for the curves with $N=512$, $L=512$, $\gamma=2.5$ and with the values of $\alpha=0.0, 5.0\times10^{-4}, 5.0\times10^{-3}$. Regression fits with $P(\tau)\sim\tau^{-\delta}$ gives the value of $\delta=1.67$ for all three values of $\alpha$. In (c) the values of $\delta$ corresponding to different values of $N(=L)$ resulting from regression fits are provided in legend.}
\label{fig-6}
\end{figure}

When a forager comes across a site $(x,y)$ having targets, the time spent at that site, before it moves away, is in general equal to the number of targets present, $h(x,y)$ (unless other foragers are also present at the site). Large values of $\gamma$, result in sites with $h(x,y)>>1$. As a result the presence of a forager at such a site is of longer duration. Now, if $\alpha$ is also small then such a forager, at any instant, draws towards the site a large number of foragers who execute TWs. During the time the former spends consuming targets, the TWs of the latter align. Such an alignment of the paths of a set of moving foragers generates an overall pattern of a moving column. Such a case is illustrated in Fig.~\ref{fig-4}(b) where two such perpendicular columns are seen giving rise to a wedge. Such a formation persists and as a whole travels until the adjacent targets are not depleted.


The observation of different patterns encouraged us to investigate the general nature of collective motion. We measured the order parameter popularly used to characterize motion of flocks of self-propelled particles \cite{vicsek}. As the motion of the foragers is restricted to the lattice, the order parameter is
\begin{equation}
\Phi=\left\langle{\sqrt{(f_{x+}-f_{x-})^2+(f_{y+}-f_{y-})^2}}\right\rangle,
\end{equation}  
where $f_{x+}$ denotes the fraction of foragers moving towards the positive $x$-direction at any instant and likewise. Temporal and configuration averaging is denoted by $\langle\cdots\rangle$ and is similar to the calculation of $\eta$. In Fig.~\ref{fig-3}(b) the plot $\Phi$ versus $\alpha$ is shown. We find that the maximum of $\eta$ is accompanied by a transition from a regime where the value of $\Phi$ is dominated by fluctuations due to random movement of the foragers to a higher values of $\Phi$. The higher values of $\Phi$ occur when foragers migrate from one cluster to another. However, the actual values of the order parameter are low because of the movement on the lattice and that just after the depletion of a cluster, the forager diffuse in all directions.

The influence of the collective searching on the movement of individual foragers is evidenced in the distribution of lengths of TWs. The tendency to search independently is less when $\alpha$ is small and long TWs are probable. In Fig.~\ref{fig-5} we plot the probability distribution of $P(l)$ of length of TWs of length $l$. We find that TWs become scale-free in the limit of $\alpha\to0$ for patchy distributions. For $N=512$, $L=512$, $\gamma=2.5$ and $\alpha=0.0$ we find $P(l)\sim l^{-\mu}$ with $\mu=-2.80$ (using regression analysis \cite{grace}). As $\alpha$ increases the TWs become shorter and the distribution falls off faster. These are shown in Fig.~\ref{fig-5}(a). For random distribution of targets and extremely clustered distribution ($\gamma=3.5$) which correspond to moving bands and columns, respectively, there is a deviation from power-law (Fig.~\ref{fig-5}(b)). In the limit $\alpha\to0$ we generally find that scale-free TWs arise for values of $N$ which are either equal to $L$ or larger but close to $L$ (Fig.~\ref{fig-5}(c) and (d)). For $N\gg L$ the distribution is fat-tailed but not a power-law. In case of $N<L$ the clusters are exploited so fast that long TWs become rare.  

The interdependence is also evident in the temporal behaviour of model. When targets are clustered there is usually a waiting time corresponding to the searching phase between the discovery of clusters.  We define $\tau$ as the time between successive encounters with targets by the group as a whole. In Fig.~\ref{fig-6} we plot the probability distribution $P(\tau)$. With the decrease in the value of $\alpha$ the states of the foragers become increasingly correlated. This is reflected in the power-law nature of $P(\tau)$. For $N=512=L$ and $\gamma=2.5$ we find $P(\tau)\sim\tau^{-\delta}$ with $\delta=-1.67$ for values of $\alpha$ between $5.0\times 10^{-3}$ to $0.0$ (Fig~\ref{fig-6}(a)). We find the power-law form for $P(\tau)$ to be retained for larger values of $\gamma$ as well. The value of $\gamma$ becoming lesser finally leading to a random distribution of targets makes $P(\tau)$ fall faster (Fig~\ref{fig-6}(b)). Faster decay of $P(\tau)$ also occurs when $N>L$ (Fig.~\ref{fig-6}(c) and (d)). This is because when both targets as well as foragers are abundant the long waiting times become rear.    

\section{Conclusion}
\label{sum}
Our model shows that the presence of interactions within a group of foragers can help maximize the efficiency of searching. The decision of an individual as well as the cost is assumed to depend on distances between members. The optimal strategy, from which the group is benefited as a whole, is a mixture of searching independently and joining other members. The dependence of the efficiency on the parameter $\alpha$ and the fact that the $\alpha$ is the inverse of a length scale also supports the notion that the exchange of information above a length scale may not be really useful for the group as a whole \cite{gazelle}. 

We also find that such interactions can lead to scale-free walks by the foragers in certain limits. Interdependence between foragers also correlates the states of the foragers which manifest in power-law distributed waiting times. The model also shows that under certain conditions stable and transient patterns of collective motion can emerge. A natural extension of the model would be to consider L\'evy walks for motion of foragers searching independently. It would be also interesting to incorporate flock cohesion in the model and study its influence on the efficiency of foraging. We believe that our model can help understand overall foraging patterns of animal groups in terms of simple strategies at the level of the individual animals. In addition it can provide guiding principles in design of artificial foraging swarms \cite{robot}. 

\begin{acknowledgments}
We acknowledge the support from the FP7 ERC COLLMOT grant. K.B. acknowledges the support from the BITS Research Initiation Grant Fund. K.B. is also thankful to A.K. Nandi for helpful discussions during the course of this research. 
\end{acknowledgments}


\begin{thebibliography}{99}
\bibitem{levy-opt}
Viswanathan, G.~M., Buldyrev, S.~V., Havlin, S., da Luz, M.~G.~E., Raposo, E.~P., Stanley, H.~E. 1999 Optimizing the success of random searches. {\it Nature}  {\bf 401}, 911-914.

\bibitem{oft}
Bartumeus, F., Catalan, J. 2009 Optimal search behavior and classic foraging theory. {\it J. Phys. A: Math. Theor.} {\bf 42}, 434002.

\bibitem{fish}
Pitcher, T. J., Magurran, A. E., Winfield, I. J. 1982 Fish in larger shoals find food faster. {\it Behav Ecol Sociobiol} {\bf 10}, 149-151.

\bibitem{sea-bird}
Haney, J. C., Fristrup, K. M., Lee, D. S. 1992 Geometry of visual recruitment by seabirds to ephemeral foraging flocks. {\it Ornis Scandinavica} {\bf 23}, 49-62.

\bibitem{optimal}
Giraldeau, L. A., Beauchamp, G. 1999 Food exploitation: searching for the optimal joining policy. {\it Trends Ecol Evol.} {\bf 14}, 102-106.

\bibitem{vickery}
Vickery, W. L., Giraldeau, L. A., Templeton, J. J., Kramer, D. L., Chapman, C. A. 1991 Producers, scroungers, and group foraging. {\it Am. Nat.} {\bf 137}, 847-863.

\bibitem{levy-data}
Viswanathan, G. M., Afanasyev, V., Buldyrev, S.~V., Murphy, E.~J., Prince, P.~A., Stanley, H.~E. 1996 L\'evy flight search patterns of wandering albatrosses. {\it Nature} {\bf 381}, 413-415.

\bibitem{sims}
Sims, D. W. {\it et al}. 2008 Scaling laws of marine predator search behaviour. {\it Nature} {\bf 451}, 1098-1102.

\bibitem{foraging-book-1}
Viswanathan, G. M., da Luz, M. G. E., Raposo, E. P., Stanley, H. E. 2011 {\it The physics of foraging}. Cambridge, UK: Cambridge University Press.

\bibitem{beauchamp-main}
Beauchamp, G. 2008 A spatial model of producing and scrounging. {\it Anim. Behav.} {\bf 76}, 1935-1942.

\bibitem{tania}
Tania, N., Vanderlei, B., Heath, J. P., Edelstein-Keshet, L. 2012 Role of social interactions in dynamic patterns of resource patches and forager aggregation. {\it Proc. Natl. Acad. Sci.} {\bf 109}, 11228-11233.

\bibitem{couzin}
Torney, C. J., Berdahl, A., Couzin, I. D. 2011 Signalling and the evolution of cooperative foraging in dynamic environments. {\it PLoS Comput Biol} {\bf 7}, e1002194.

\bibitem{spider-monkey-exp}
Ramos-Fern\'andez, G., Mateos, J. L., Miramontes, O., Cocho, G., Larralde, H., Ayala-Orozco, B. 2004 L\'evy walk patterns in the foraging movements of spider monkeys (Ateles geoffroyi). {\it Behav Ecol Sociobiol} {\bf 55}, 223-230.

\bibitem{monkey-model}
Boyer, D., Miramontes, O., Ramos-Fern\'andez, G., Mateos, J. L., Cocho, G. 2004 Modeling the searching behavior of social monkeys. {\it Physica A} {\bf 342}, 329-335.

\bibitem{zhou}
Han, X.-P., Zhou, T., Wang, B.~H. 2011  Scaling mobility patterns and collective movements: Deterministic walks in lattices. {\it Phys. Rev. E} {\bf 83},  056108.

\bibitem{honeybee}
de Vries, H., Biesmeijer, J. C. 1998 Modelling collective foraging by means of individual behaviour rules in honey-bees. {\it Behav Ecol Sociobiol}, {\bf 44}, 109-124.

\bibitem{honeybee-model}
Dechaume-Moncharmont, F. X., Dornhaus, A., Houston, A. I., McNamara, J. M., Collins, E. J., Franks, N. R. 2005 The hidden cost of information in collective foraging. {\it Proc Biol Sci} {\bf 272}, 1689-1695.

\bibitem{ant}
Cerd\'a, X., Angulo, E., Boulay, R., Lenoir, A. 2009 Individual and collective foraging decisions: a field study of worker recruitment in the gypsy ant Aphaenogaster senilis. {\it Behav Ecol Sociobiol} {\bf 63}, 551-562.

\bibitem{gazelle}
Mart\'{\i}nez-Garc\'{\i}a, R., Calabrese, J. M., Mueller, T., Olson, K. A., L\'opez, C. 2013 Optimizing the search for resources by sharing information: Mongolian gazelles as a case study. {Phys. Rev. Lett.} {\bf 110}, 248106. 


\bibitem{patch-2}
P\"oys\"a, H. 1992 Group foraging in patchy environments: the importance of coarse-level local enhancement. {\it Ornis Scandinavica} {\bf 23} 159-166.

\bibitem{monkey-model-2}
Boyer, D., Ramos-Fern\'andez, G., Miramontes, O., Mateos, J. L., Cocho, G., Larralde, H., Ramos, H., Rojas, F. 2006 Scale-free foraging by primates emerges from their interaction with a complex environment. {\it Proc. R. Soc. Lond. B} {\bf 273}, 1743–1750.

\bibitem{hierarchical-patch}
Fauchald, P. 1999 Foraging in a hierarchical patch system. {\it Am. Nat.} {\bf 153}, 603-613.

\bibitem{cost1}
MacArthur, R. H., Pianka, E. R. 1966 On optimal use of a patchy environment. {\it Am. Nat.} {\bf 100}, 603-609.

\bibitem{cost2}
Ruxton, G. D., Fraser, C., Broom, M. 2005 An evolutionarily stable joining policy for group foragers. {\it Behav Ecol} {\bf 16}, 856-864.

\bibitem{gurney}
Ruxton, G. D., Hall, S. J., Gurney, W. S. C. 1995 Attraction toward feeding conspecifics when food patches are exhaustible. {\it Am. Nat.} {\bf 145}, 653-660.

\bibitem{vicsek}
Vicsek, T., Czir\'ok, A., Ben-Jacob, E., Cohen, I., Shochet, O. 1995 Novel type of phase transition in a system of self-driven particles. {\it Phys. Rev. Lett.} {\bf 75}, 1226.

\bibitem{grace}
Grace (Version 5.1.23) [Software]. Available from \url{http://plasma-gate.weizmann.ac.il/Grace/}

\bibitem{robot}
Dai, H. 2010 {\it Design of adaptive collective foraging in swarm robotic systems}. Kalamazoo, MI, USA: Western Michigan University.

\end{thebibliography}
\end{document}